\documentclass[a4paper]{amsart}
\usepackage{color}
\usepackage[latin1]{inputenc}
\usepackage[T1]{fontenc}
\usepackage{amsfonts}
\usepackage{amssymb}
\usepackage{amsmath}
\usepackage{amsthm}
\usepackage{graphicx,type1cm,eso-pic,color}
\usepackage{pstricks,pst-plot,pstricks-add}
\usepackage{float}
\newtheorem{theorem}{Theorem}[section]

\newtheorem{proposition}[theorem]{Proposition}
\newtheorem{corollary}[theorem]{Corollary}
\newtheorem{definition}[theorem]{Definition}
\newtheorem{assumption}[theorem]{Assumption}
\begin{document}
\setlength\arraycolsep{2pt}
\title{A skew normal model of the dose-effect relation in pharmacology}
\author{Francis LAVERGNE}
\author{Nicolas MARIE}
\address{Medical Trial, Paris, France}
\email{lavergne.francis100@gmail.com}
\address{Laboratoire ISTI, ESME Sudria, Paris, France}
\email{marie@esme.fr}
\date{}
\maketitle
%


%
\begin{abstract}
This paper deals with a skew-normal model of the relation between a dose $d > 0$ and a quantitative measure $\textrm E(d)$ of an effect of the administered drug. Precisely, $\textrm E(d)$ is a measure of the therapeutic response or a measure of a side-effect. Some existing and additional properties of the logistic functions are proved, and a skew-normal model of the escape time of rats under an experimental antidepressant medication is provided.
\end{abstract}
\tableofcontents
\noindent
\textbf{Acknowledgements.} Many thanks to Th\'er\`ese M. Jay for the psychopharmacological datas used at Section 4.
%


%
\section{Introduction}
This paper deals with a skew-normal model of the relation between a dose $d > 0$ and a quantitative measure $\textrm E(d)$ of an effect of the administered drug. Precisely, $\textrm E(d)$ is a measure of the therapeutic response or a measure of a side-effect.
\\
\\
As explained in Prentice \cite{PRENTICE76} and Brown \cite{BROWN78}, it is usual to put
\begin{displaymath}
\textrm E(d) :=\int_{0}^{(d -\mu)/\sigma}f(x)dx
\end{displaymath}
where, $\mu\in\mathbb R$, $\sigma > 0$ and $f :\mathbb R\rightarrow\mathbb R$ is a probability density function. Often, $f$ is a probit or a logit function. The parameters $\mu$ and $\sigma$ are estimated from the observations. For an application on clinical datas, for instance, see Verlato et al. \cite{VERLATO96}.
\\
\\
The skew-normal distribution has been already used to model the quantal response. In \cite{WAGNER07}, Section 3.5, Wagner assumes that $f$ is a skew-normal density function. On the skew-normal distribution, see Azzalini \cite{AZZALINI85} and Chen et al. \cite{CGN04}. The basics on the skew-normal distribution are stated at Section 3.1. On the multidimensional skew-normal distributions and an application in neurotoxicology, see T. Baghfalaki et al. \cite{BGK12}.
\\
\\
The model studied in this paper is a family of random variables $\textrm E :=\{\textrm E(d), d\in\mathbb R_+\}$ satisfying the following conditions :
\begin{enumerate}
 \item For every $d\in\mathbb R_+$, $\textrm E(d)\rightsquigarrow\mathcal N(\xi(d),\omega^2(d),\alpha(d))$ where $\xi,\alpha\in\textrm C^0(\mathbb R_+,\mathbb R)$ and $\omega\in\textrm C^0(\mathbb R_+,]0,\infty[)$.
 \item $\mu : d\in\mathbb R_+\longmapsto\mathbb E[\textrm E(d)]$ is a logistic function.
 \item There exists $d_0\in\mathbb R_+$ such that $\sigma : d\in\mathbb R_+\longmapsto\sqrt{\textrm{var}[\textrm E(d)]}$ is decreasing on $[d_0,\infty[$, and
 \begin{displaymath}
 \lim_{d\rightarrow\infty}\sigma(d) = 0.
 \end{displaymath}
\end{enumerate}
Assume that $\textrm E(d)$ is a measure of the therapeutic response of the administered drug at the dose $d\in\mathbb R_+$. It is usual to assume that $\textrm E(d)$ is a Gaussian random variable. The skewness of the empirical distributions are taken into account in the model $\textrm E$ of the therapeutic response in order to refine the choice of an optimal dose. Indeed, the skewness of the distribution of $\textrm E(d)$ indicates if the therapeutic responses of the major part of the patients are over (positive skewness coefficient) or under (negative skewness coefficient) the mean therapeutic response. Therefore, if the mean of $\textrm E(d)$ is high enough, its skewness coefficient is positive and its standard deviation is small enough, then the dose $d$ is admissible. In fact, ideally, the optimal dose should maximise the mean and the skewness coefficient, and minimize the standard deviation of the therapeutic response.
\\
\\
Section 2 deals with some existing and additional properties of the logistic functions. Section 2 provides an approximation method of the parameters of the logistic functions in the most general case, which is used at Section 4. The proofs of the results stated at Section 2 are detailed at Appendix A.
\\
\\
Section 3 deals with the estimation of the functions $\mu$ and $\sigma$, and then of the functions $\xi$, $\omega$ and $\alpha$. So, for each admissible dose $d > 0$, the model $\textrm E$ allows to simulate the measure of the effect $\textrm E(d)$ multiple times for a better evaluation at the dose $d$. As mentioned above, the basics on the skew-normal distribution are stated at Section 3.1.
\\
\\
Section 4 deals with a model of the escape time of rats under an experimental antidepressant medication. In psychopharmacology, since the medications are often administered for several months or years, it is crucial to find the smallest efficient dose. See Lavergne and Jay \cite{LJ10} about the efficiency of antidepressant medications for small doses.
\\
\\
In \cite{HS81}, Holford and Sheiner studied the relationship between the therapeutic response and the elimination process of the drug. In a forthcoming work, the dose-effect model studied in this paper will be related to the fractional pharmacokinetics model studied in Marie \cite{MARIE14}. 
%


%
\section{Approximation of the parameters of the logistic functions}
This section deals with some existing and additional properties of the logistic functions.
\\
\\
Consider $n\in\mathbb N^*$ observations $y_1,\dots,y_n > 0$ of a logistic function $f$ at $x_1,\dots,x_n$ respectively. Many authors approximate the parameters of the function $f$ by linear regression on $(x_1,z_1),\dots,(x_n,z_n)$, where
\begin{displaymath}
z_i :=
\log\left[\left(
\frac{y_i}{y^*}\right)^{-1} - 1\right]
\textrm{ $;$ }
\forall i\in\{1,\dots,n\}
\end{displaymath}
and
\begin{displaymath}
y^* :=
\max_{i\in\{1,\dots,n\}}y_i.
\end{displaymath}
The major drawback of that method is to assume that
\begin{displaymath}
\lim_{x\rightarrow -\infty}f(x) = y^*
\textrm{ or }
\lim_{x\rightarrow\infty}f(x) = y^*.
\end{displaymath}
This section provides an approximation method of the parameters of the logistic functions in the most general case, which is used at Section 4. The proofs of the results stated in this section are detailed at Appendix A.
%


%
\begin{definition}\label{logistic_functions}
$f :\mathbb R\rightarrow\mathbb R$ is a logistic function if and only if,
\begin{displaymath}
f(x) := l_1 +\frac{1}{(l_2 - l_1)^{-1} + e^{mx + p}}
\textrm{ $;$ }
\forall x\in\mathbb R
\end{displaymath}
with $m\in\mathbb R^*$, $p\in\mathbb R$, and $l_1,l_2\in\mathbb R$ such that $l_2 > l_1$.
\end{definition}
Throughout this section, $f$ is the logistic function defined at Definition \ref{logistic_functions}.
%


%
\begin{proposition}\label{logistic_behaviour}
The logistic function $f$ satisfies the following properties :
\begin{enumerate}
 \item If $m < 0$ (resp. $m > 0$), then
 \begin{displaymath}
 \lim_{x\rightarrow -\infty}f(x) = l_1
 \textrm{ and }
 \lim_{x\rightarrow\infty}f(x) = l_2
 \textrm{ (resp.}
 \lim_{x\rightarrow -\infty}f(x) = l_2
 \textrm{ and }
 \lim_{x\rightarrow\infty}f(x) = l_1
 \textrm{).}
 \end{displaymath}
 \item If $m < 0$ (resp. $m > 0$), then $f$ is increasing (resp. decreasing) on $\mathbb R$.
 \item The graph of the function $f$ has a unique inflection point, at
 \begin{displaymath}
 \theta := -\frac{\log(l_2 - l_1) + p}{m}.
 \end{displaymath}
\end{enumerate}
\end{proposition}
%


%
\begin{corollary}\label{mprl_expressions}
The parameter $l_1$ is a solution of the following equation :
\begin{equation}\label{equation_l}
\log\left[
2\frac{f(\theta) - f(0)}{f(0) - l_1} + 1\right] -
\frac{2\theta f'(\theta)}{f(\theta) - l_1} = 0.
\end{equation}
Moreover,
\begin{displaymath}
l_2 = 2f(\theta) - l_1
\textrm{, }
m = -\frac{2f'(\theta)}{f(\theta) - l_1}
\textrm{ and }
p =\log\left[\frac{1}{f(0) - l_1} -\frac{1}{2[f(\theta) - l_1]}\right].
\end{displaymath}
\end{corollary}
%


%
\begin{proposition}\label{logistic_ODE}
The logistic function $f$ is the solution of the following ordinary differential equation :
\begin{equation}\label{logistic_equation}
y(x) =
f(0) -
m\int_{0}^{x}
[y(u) - l_1]\left[
1 -\frac{1}{l_2 - l_1}[y(u) - l_1]\right]du.
\end{equation}
\end{proposition}
Consider $n\in\mathbb N^*$ observations $y_1,\dots,y_n$ of the logistic function $f$ at $x_1 = 0,x_2\dots,x_n$ respectively. The end of the section is devoted to some methods to approximate the parameters $m$, $p$, $l_1$ and $l_2$ of the logistic function $f$.
\\
\\
On one hand, assume that the values of the parameters $l_1$ and $l_2$ are known. Let $\Phi : ]l_1,l_2[\rightarrow\mathbb R$ be the map defined by :
\begin{displaymath}
\Phi(x) :=
\log\left(\frac{1}{x - l_1} -\frac{1}{l_2 - l_1}\right)
\textrm{ $;$ }
\forall x\in ]l_1,l_2[.
\end{displaymath}
By Proposition \ref{logistic_ODE} together with the change of variable formula for the Riemann-Stielj\`es integral :
\begin{eqnarray*}
 \Phi[f(x)] & = &
 \Phi[f(0)] +
 \int_{0}^{x}\Phi'[f(u)]df(u)\\
 & = &
 \Phi[f(0)] -\\
 & &
 m\int_{0}^{x}\Phi'[f(u)]
 [f(u) - l_1]\left[
 1 -\frac{1}{l_2 - l_1}[f(u) - l_1]\right]du\\
 & = &
 p - mx.
\end{eqnarray*}
So, if $l_1$ and $l_2$ are known, the parameters $m$ and $p$ can be approximated by linear regression on $(x_1,z_1),\dots,(x_n,z_n)$, where
\begin{displaymath}
z_i :=\Phi(y_i)
\textrm{ ; }
\forall i\in\{1,\dots,n\}.
\end{displaymath}
Precisely,
\begin{eqnarray*}
 \widehat m_n & := &
 -\frac{\displaystyle{-\frac{1}{n - 1}\left(\sum_{i = 1}^{n}x_i\right)\left(\sum_{i = 1}^{n}z_i\right) +\sum_{i = 1}^{n}x_iz_i}}
 {\displaystyle{-\frac{1}{n - 1}\left(\sum_{i = 1}^{n}x_i\right)^2 +\sum_{i = 1}^{n}x_{i}^{2}}}
 \textrm{ and}\\
 \widehat p_n & := &
 \frac{1}{n - 1}\sum_{i = 1}^{n}z_i +
 \frac{\widehat m_n}{n - 1}\sum_{i = 1}^{n}x_i
\end{eqnarray*}
are some unbiased estimators of $m$ and $p$ respectively. This method is classic (see R.F. Gunst and R.L. Mason \cite{GM80}).
\\
\\
On the other hand, assume that the values of the parameters $l_1$ or $l_2$ are unknown. In that case, the transformation $\Phi$ cannot by applied to $y_1,\dots,y_n$. In the current paper, an alternative method of approximation is provided.
\\
\\
Since $\textrm{card}(\{1,\dots,n - 1\}) <\infty$, the maximization problem
\begin{displaymath}
\max_{i\in\{1,\dots,n - 1\}}
\left|\frac{y_{i + 1} - y_i}{x_{i + 1} - x_i}\right|
\end{displaymath}
has a unique solution $n(\theta)$. So,
\begin{displaymath}
\theta_n :=
\frac{x_{n(\theta) + 1} + x_{n(\theta)}}{2}
\textrm{, }
\gamma_n =
\frac{y_{n(\theta) + 1} + y_{n(\theta)}}{2}
\textrm{ and }
\delta_n :=
\frac{y_{n(\theta) + 1} - y_{n(\theta)}}{x_{n(\theta) + 1} - x_{n(\theta)}}
\end{displaymath}
define some converging approximations of $\theta$, $f(\theta)$ and $f'(\theta)$ respectively.
\\
\\
If the value of $l_1$ is known but not the value of $l_2$, then by Corollary \ref{mprl_expressions} :
\begin{displaymath}
l_2(n) :=
2\gamma_n - l_1
\end{displaymath}
defines an approximation of $l_2$. So, the previous method is adaptable by replacing $\Phi$ by the map $\Phi_n : ]l_1,l_2(n)[\rightarrow\mathbb R$ defined by :
\begin{displaymath}
\Phi_n(x) :=
\log\left[\frac{1}{x - l_1} -\frac{1}{l_2(n) - l_1}\right]
\textrm{ $;$ }
\forall x\in ]l_1,l_2(n)[.
\end{displaymath}
If the values of $l_1$ and $l_2$ are both unknown, let $l_1(n)$ be the numerical approximation of the solution of the following equation :
\begin{equation}\label{equation_l_approximation}
\frac{\gamma_n - y_1}{y_1 - l_1(n)} +\frac{1}{2} -
\frac{1}{2}\exp\left[\frac{2\theta_n\delta_n}{\gamma_n - l_1(n)}\right] = 0.
\end{equation}
It defines an approximation of $l_1$ because Equation (\ref{equation_l_approximation}) is a discretization of Equation (\ref{equation_l}). Moreover, by Corollary \ref{mprl_expressions} :
\begin{displaymath}
l_2(n) :=
2\gamma_n - l_1(n)
\textrm{, }
m_n :=
-\frac{2\delta_n}{\gamma_n - l_1(n)}
\textrm{ and }
p_n :=
\log\left[\frac{1}{y_1 - l_1(n)} -\frac{1}{2[\gamma_n - l_1(n)]}\right]
\end{displaymath}
define some converging approximations of $l_2$, $m$ and $p$ respectively.
%


%
\section{A model of the dose-effect relation}
The model introduced in this section is tailor-made to study the relation between a dose $d > 0$ and a quantitive measure $\textrm E(d)$ of an effect of the administered drug. Precisely, $\textrm E(d)$ can be a measure of the therapeutic response or a measure of a side-effect.
%


%
\subsection{The skew normal distribution}
On the skew normal distribution, see Azzalini \cite{AZZALINI85} and Chen et al. \cite{CGN04}.
%


%
\begin{definition}\label{asymmetric_Gaussian}
The skew normal distribution of parameters $\xi\in\mathbb R$ (location), $\omega > 0$ (scale) and $\alpha\in\mathbb R$ (shape) is the probability measure $\Pi$ on $(\mathbb R,\mathcal B(\mathbb R))$ defined by :
\begin{displaymath}
\Pi(dx) =
\frac{dx}{\omega\sqrt{2\pi}}
\exp\left(-\frac{|x -\xi|^2}{2\omega^2}\right)
\left[1 +\normalfont{\textrm{erf}}\left[
\frac{\alpha(x -\xi)}{\omega\sqrt 2}
\right]\right].
\end{displaymath}
The skew normal distribution of parameters $\xi$, $\omega$ and $\alpha$ is denoted by $\mathcal N(\xi,\omega^2,\alpha)$.
\end{definition}
Consider $\xi,\alpha\in\mathbb R$ and $\omega > 0$. The parameters of the skew normal distribution $\mathcal N(\xi,\omega^2,\alpha)$ are related to its mean $\mu$, its standard deviation $\sigma$ and to its skewness coefficient $\gamma$ (Pearson) as follow :
\begin{displaymath}
\alpha =\frac{\delta}{\sqrt{1 -\delta^2}}
\textrm{, }
\omega^2 =\frac{\sigma^2}{1 - 2\delta^2/\pi}
\textrm{ and }
\xi =\mu -\omega\delta\sqrt{\frac{2}{\pi}}
\end{displaymath}
with
\begin{displaymath}
|\delta| :=
\frac{|\gamma|^{1/3}\sqrt{\pi/2}}{\sqrt{|\gamma|^{2/3} + [(4 -\pi)/2]^{2/3}}}.
\end{displaymath}
Let $(X_1,\dots,X_n)$ be a $n$-sample ($n\in\mathbb N^*$) such that $X_1\rightsquigarrow\mathcal N(\xi,\omega^2,\alpha)$. Consider
\begin{displaymath}
\overline X_n :=
\frac{1}{n}\sum_{i = 1}^{n}X_i
\textrm{, }
S_{n}^{2} :=
\frac{1}{n}\sum_{i = 1}^{n}(X_i -\overline X_n)^2
\textrm{ and }
\widehat\gamma_n :=
\frac{1}{n}\sum_{i = 1}^{n}
\left(\frac{X_i -\overline X_n}{S_n}\right)^3.
\end{displaymath}
So,
\begin{displaymath}
\widehat\alpha_n :=\frac{\widehat\delta_n}{\sqrt{1 -\widehat\delta_{n}^{2}}}
\textrm{, }
\widehat\omega_{n}^{2} :=\frac{S_{n}^{2}}{1 - 2\widehat\delta_{n}^{2}/\pi}
\textrm{ and }
\widehat\xi_n :=\overline X_n -\widehat\omega_n\widehat\delta_n\sqrt{\frac{2}{\pi}}
\end{displaymath}
with
\begin{displaymath}
|\widehat\delta_n| :=
\frac{|\widehat\gamma_n|^{1/3}\sqrt{\pi/2}}{\sqrt{|\widehat\gamma_n|^{2/3} + [(4 -\pi)/2]^{2/3}}}
\end{displaymath}
are some converging estimators of $\alpha$, $\omega^2$ and $\xi$ respectively.
%


%
\subsection{The skew normal model of the D-E relation}
The model is a family of random variables $\textrm E :=\{\textrm E(d),d\in\mathbb R_+\}$ satisfying the following assumption :
\begin{assumption}\label{model_assumptions}
\white .\black
\begin{enumerate}
 \item For every $d\in\mathbb R_+$, $\normalfont{\textrm E}(d)\rightsquigarrow\mathcal N(\xi(d),\omega^2(d),\alpha(d))$ where $\xi,\alpha\in\normalfont{\textrm C}^0(\mathbb R_+,\mathbb R)$ and $\omega\in\normalfont{\textrm C}^0(\mathbb R_+,]0,\infty[)$.
 \item $\mu : d\in\mathbb R_+\longmapsto\mathbb E[\normalfont{\textrm E(d)}]$ is a logistic function.
 \item There exists $d_0\in\mathbb R_+$ such that $\sigma : d\in\mathbb R_+\longmapsto\sqrt{\normalfont{\textrm{var}}[\normalfont{\textrm E(d)}]}$ is decreasing on $[d_0,\infty[$, and
 \begin{displaymath}
 \lim_{d\rightarrow\infty}\sigma(d) = 0.
 \end{displaymath}
\end{enumerate}
\end{assumption}
Assumption \ref{model_assumptions}.(3) means that the variability of the measured effect between the patients tends to vanish for large doses of the drug.
\\
\\
Assume that $\textrm E(d)$ is a measure of the therapeutic response of the administered drug at the dose $d\in\mathbb R_+$. It is usual to assume that $\textrm E(d)$ is a Gaussian random variable. The skewness of the empirical distributions are taken into account in the model $\textrm E$ of the therapeutic response in order to refine the choice of an optimal dose. Indeed, the skewness of the distribution of $\textrm E(d)$ indicates if the therapeutic responses of the major part of the patients are over (positive skewness coefficient) or under (negative skewness coefficient) the mean therapeutic response. Therefore, if the mean of $\textrm E(d)$ is high enough, its skewness coefficient is positive and its standard deviation is small enough, then the dose $d$ is admissible.
\\
\\
With the notations of Assumption \ref{model_assumptions}, the last part of the section deals with the estimation of the functions $\mu$ and $\sigma$, and then of the functions $\xi$, $\omega$ and $\alpha$. So, for each admissible dose $d > 0$, Assumption \ref{model_assumptions}.(1) allows to simulate the measure of the therapeutic response $\textrm E(d)$ multiple times for a better evaluation at the dose $d$.
\\
\\
Consider the finite set $\textrm D\subset\mathbb R_+$ of the administered doses and, for every $d\in\textrm D$, $n\in\mathbb N^*$ independent observations $e_1(d),\dots,e_n(d)$ of the random variable $\textrm E(d)$. For each dose $d\in\textrm D$, $m(d)$, $\sigma(d)$ and $\gamma(d)$ can be estimated by
\begin{eqnarray*}
 \widehat m_n(d) & := &
 \frac{1}{n}\sum_{i = 1}^{n}e_i(d)
 \textrm{, }\\
 \widehat\sigma_{n}^{2}(d) & := &
 \frac{1}{n}\sum_{i = 1}^{n}[e_i(d) -\widehat m_n(d)]^2
 \textrm{ and}\\
 \widehat\gamma_n(d) & := &
 \frac{1}{n}\sum_{i = 1}^{n}
 \left[\frac{e_i(d) -\widehat m_n(d)}{\widehat\sigma_n(d)}\right]^3
\end{eqnarray*}
respectively.
\\
\\
Since $m$ is a logistic function by Assumption \ref{model_assumptions}, its parameters can be approximated by the method stated at Section 2.
\\
\\
Assume there exists $\widehat d_0\in\textrm D$ such that
\begin{displaymath}
d\in\textrm D
\longmapsto
\widehat\sigma_n(d)
\end{displaymath}
is decreasing on $\textrm D\cap [\widehat d_0,\infty[$.
\begin{enumerate}
 \item If the function $d\in\textrm D\longmapsto\widehat\sigma_n(d)$ is constant on $\textrm D\cap [0,\widehat d_0[$, then $\sigma$ could be a logistic function. Its parameters can be approximated by the method stated at Section 2.
 \item If the function $d\in\textrm D\longmapsto\widehat\sigma_n(d)$ is increasing on $\textrm D\cap [0,\widehat d_0[$, then $\sigma$ could be a Gaussian-type function :
 \begin{displaymath}
 \sigma(d) := l +\exp(-md^2 + pd + q)
 \textrm{ $;$ }
 \forall d\in\mathbb R_+
 \end{displaymath}
 with $m > 0$ and $l,p,q\in\mathbb R$. Since
 \begin{displaymath}
 \lim_{d\rightarrow\infty}\sigma(d) = 0
 \end{displaymath}
 by Assumption \ref{model_assumptions}.(4), then $l = 0$. So, the parameters $m$, $p$ and $q$ can be approximated by polynomial regression on $\{(d,y_n(d)) ; d\in\textrm D\}$, where
\begin{displaymath}
y_n(d) :=\log[\widehat\sigma_n(d)]
\textrm{ ; }
\forall d\in\textrm D.
\end{displaymath}
This method is classic (see P. Armitage et al. \cite{ABM08}).
\end{enumerate}
%


%
\section{An application to an antidepressant medication}
This section deals with a model of the escape time (ET) of rats under an experimental antidepressant (AD) medication. For confidentiality reasons, the name of the drug is not specified.
\\
\\
The escape time (in seconds) is interpreted as a therapeutic response to the AD. The AD is administered at four different doses : 0, 0.75, 1.5 and 3 mg.kg$^{-1}$. The escape time is evaluated on 32 rats ; 8 rats for each posology.
\\
\\
For each dose, the mean, the standard deviation and the skewness coefficient of the escape time have been computed :
\\
\begin{center}
\begin{tabular}{l l l l l l}
\hline
Statistics | Doses & $0$ & $0.75$ & $1.5$ & $3$\\
\hline
\hline
 Mean & $33.3875$ & $44.1625$ & $51.5$ & $78.225$\\
 Standard deviation & $26.9715$ & $30.8113$ & $44.6582$ & $31.9657$\\
 Skewness & $-0.0276$ & $-0.1381$ & $1.2827$ & $0.3504$\\
\hline
\end{tabular}
\end{center}
%


%
\subsection{The model of the dose-escape time relation}
According to Assumption \ref{model_assumptions}, the mean escape time is modeled by a logistic function of the administered dose. By using the approximation procedure stated at Section 2 :
\begin{displaymath}
m_{\textrm{ET}}(d) =
21.8153 +
\frac{1}{0.0116 + e^{-0.8278\cdot d - 2.5929}}
\textrm{ $;$ }
\forall d\in\mathbb R_+.
\end{displaymath}
The standard deviation and the skewness coefficient are modeled by Gaussian type functions of the administered dose. By using polynomial regressions as suggested at Section 3 :
\begin{displaymath}
\sigma_{\textrm{ET}}(d) =
\exp(-0.1502\cdot d^2 + 0.5289\cdot d + 3.2459)
\end{displaymath}
and
\begin{displaymath}
\gamma_{\textrm{ET}}(d) =
0.2381 +
\exp(-0.6503\cdot d^2 + 2.2935\cdot d - 1.5578)
\end{displaymath}
for every $d\in\mathbb R_+$.
\\
\\
The functions $m_{\textrm{ET}}$, $\sigma_{\textrm{ET}}$ and $\gamma_{\textrm{ET}}$ are plotted on the interval of doses $[0,4]$ :
\begin{figure}[htbp]
\begin{center}
\includegraphics[scale = 0.27]{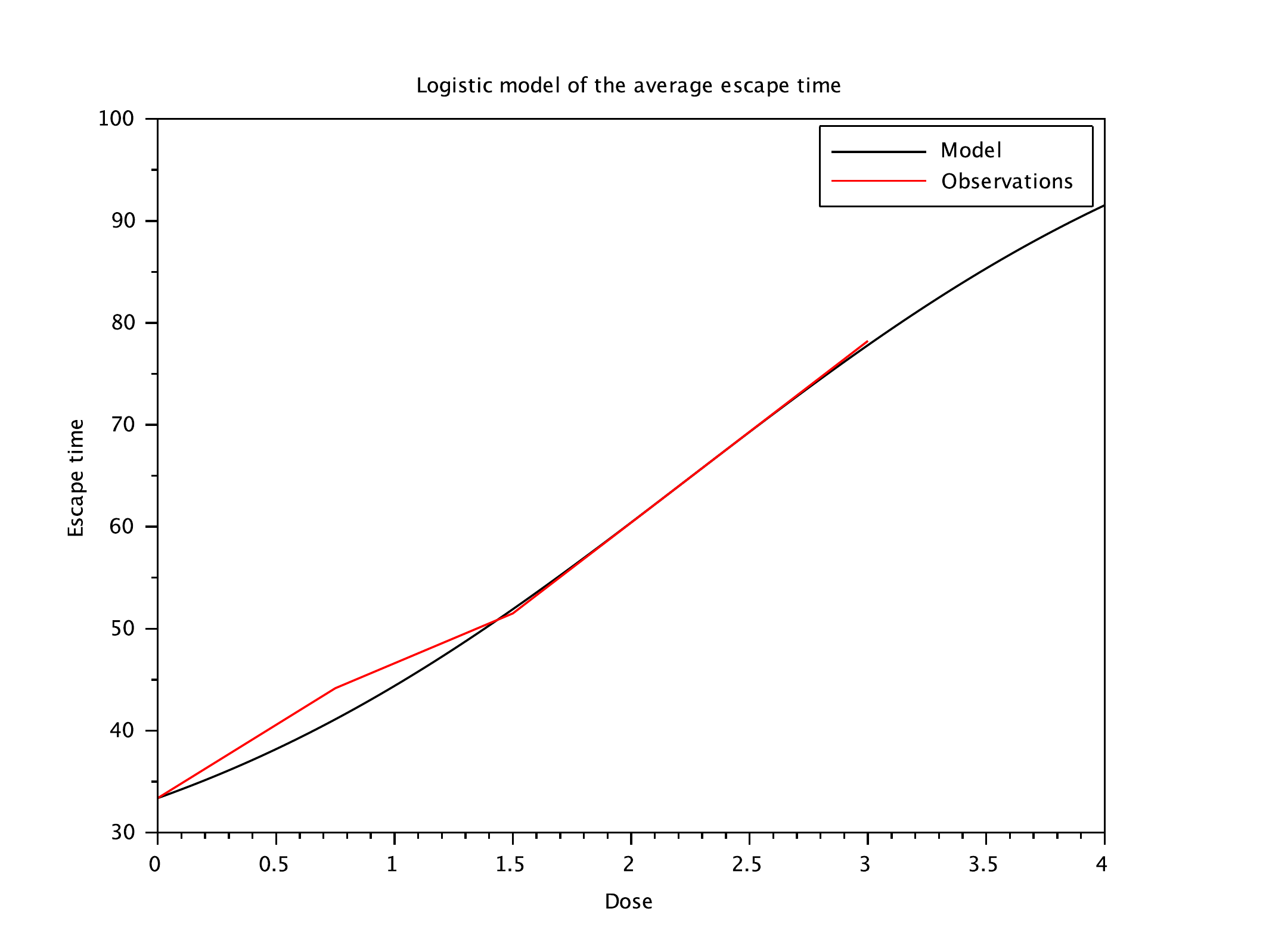}
\caption{\small{Mean ET}}
\label{fig:image1}
\end{center}
\end{figure}
\begin{figure}[htbp]
\begin{minipage}[c]{.45\linewidth}
\begin{center}
\includegraphics[scale = 0.27]{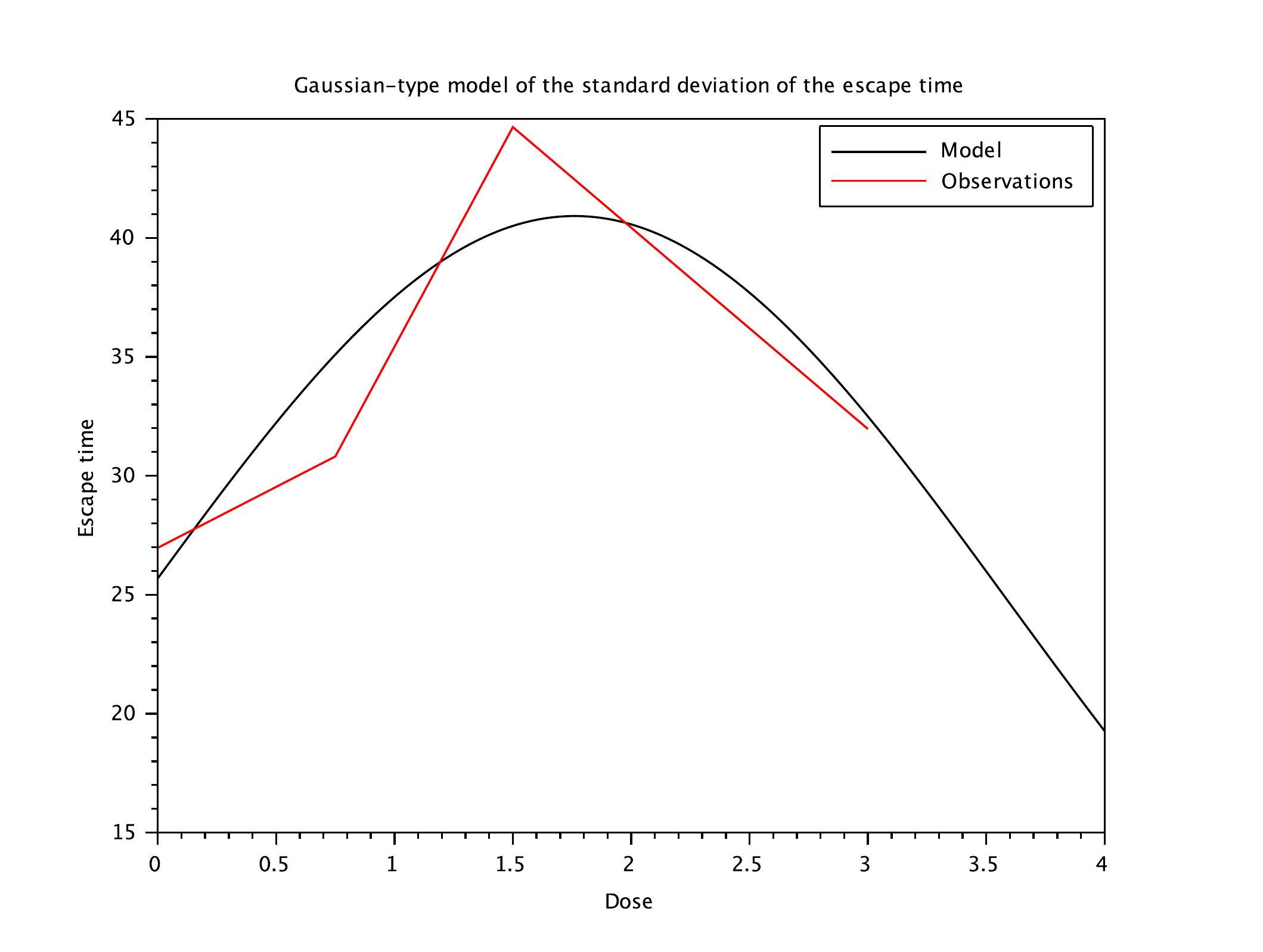}
\caption{\small{Std. dev. ET}}
\label{fig:image1}
\end{center}
\end{minipage}
\hfill
\begin{minipage}[c]{.45\linewidth}
\begin{center}
\includegraphics[scale = 0.27]{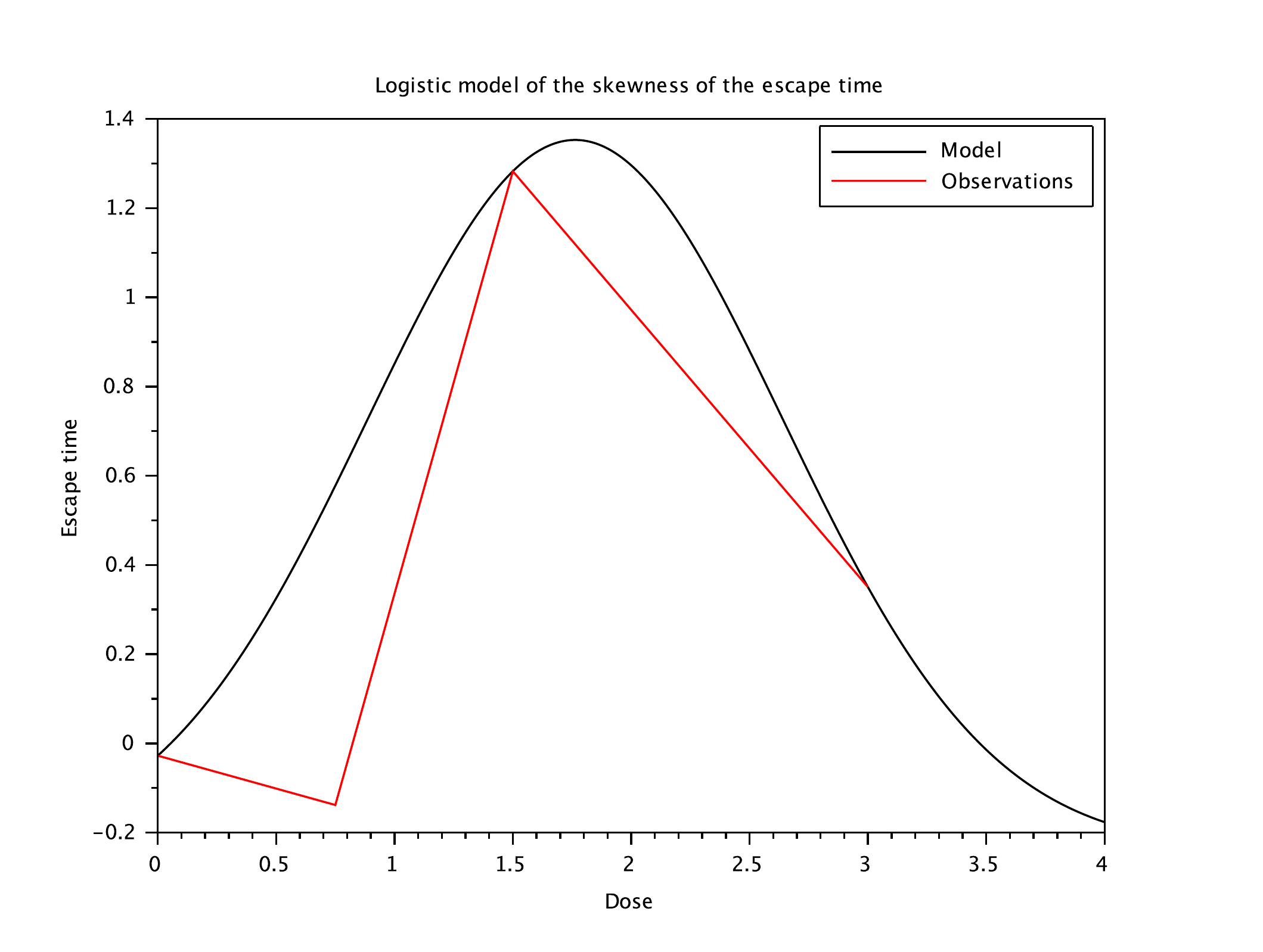}
\caption{\small{Skewness ET}}
\label{fig:image2}
\end{center}
\end{minipage}
\end{figure}
%


%
\subsection{The optimal dose}
The optimal dose should maximise the mean and the skewness coefficient, and minimize the standard deviation of the therapeutic response.
\\
\\
Unfortunately, there is no dose such that these conditions are satisfied together by the model of the dose-escape time relation.
\\
\\
Consider the dose $d_{\textrm{opt}} := 3$ which maximizes the mean on $[0,3]$, and such that $\sigma_{\textrm{ET}}(d_{\textrm{opt}}) = 32.4903$ and $\gamma_{\textrm{ET}}(d_{\textrm{opt}}) = 0.3504$.
\\
\\
Since the AD tested in the clinical trial has a priori no significant side-effect for the doses less or equal than 3 mg.kg$^{-1}$, $d_{\textrm{opt}}$ could be the optimal dose. Indeed, it maximizes the mean of the escape time, $\sigma_{\textrm{ET}}(d_{\textrm{opt}})$ is closer to
\begin{displaymath}
\min_{d\in [0,3]}\sigma_{\textrm{ET}}(d) =
26.9715
\end{displaymath}
than to
\begin{displaymath}
\max_{d\in [0,3]}\sigma_{\textrm{ET}}(d) =
44.6582,
\end{displaymath}
and $\gamma_{\textrm{ET}}(d_{\textrm{opt}}) > 0$.
%


%
\appendix
%


%
\section{The proofs of Section 2}
\noindent
\textit{Proof of Proposition \ref{logistic_behaviour}.} For every $x\in\mathbb R$,
\begin{displaymath}
f'(x) =
-\frac{me^{mx + p}}{[(l_2 - l_1)^{-1} + e^{mx + p}]^2}
\end{displaymath}
and
\begin{displaymath}
f''(x) =
-\frac{m^2e^{mx + p}}{[(l_2 - l_1)^{-1} + e^{mx + p}]^3}\left(\frac{1}{l_2 - l_1} - e^{mx + p}\right).
\end{displaymath}
\begin{enumerate}
 \item Assume that $m < 0$. So,
 \begin{displaymath}
 \lim_{x\rightarrow -\infty}f(x) = l_1 + 0 = l_1
 \end{displaymath}
 and
 \begin{displaymath}
 \lim_{x\rightarrow\infty}f(x) = l_1 +\frac{1}{(l_2 - l_1)^{-1}} = l_2.
 \end{displaymath}
 Assume that $m > 0$. So,
 \begin{displaymath}
 \lim_{x\rightarrow -\infty}f(x) = l_1 +\frac{1}{(l_2 - l_1)^{-1}} = l_2
 \end{displaymath}
 and
 \begin{displaymath}
 \lim_{x\rightarrow\infty}f(x) = l_1 + 0 = l_1.
 \end{displaymath}
 \item If $m < 0$ (resp. $m > 0$), $f'(x) > 0$ (resp. $f'(x) < 0$) for every $x\in\mathbb R$. So, if $m < 0$ (resp. $m > 0$), then $f$ is increasing (resp. decreasing) on $\mathbb R$.
 \item $f''(x) = 0$ if and only if $x =\theta$. Moreover, if $m < 0$ (resp. $m > 0$), $f''(x) > 0$ if and only if $x\in ]-\infty,\theta[$ (resp. $x\in]\theta,\infty[$). So, the graph of the function $f$ has a unique inflection point, at $\theta$.
\end{enumerate}
That achieves the proof. $\square$
\\
\\
\textit{Proof of Corollary \ref{mprl_expressions}.} By Definition \ref{logistic_functions} :
\begin{equation}\label{mprl_1}
f(0) = l_1 +\frac{1}{(l_2 - l_1)^{-1} + e^p}.
\end{equation}
By Proposition \ref{logistic_behaviour} :
\begin{eqnarray}
 \label{mprl_2}
 \theta & = &
 -\frac{\log(l_2 - l_1) + p}{m},\\
 \label{mprl_3}
 f(\theta) & = & \frac{l_1 + l_2}{2}\textrm{ and}\\
 \label{mprl_4}
 f'(\theta) & = & -\frac{m(l_2 - l_1)}{4}.
\end{eqnarray}
By Equation (\ref{mprl_3}), $l_2 = 2f(\theta) - l_1$. By Equation (\ref{mprl_4}) :
\begin{eqnarray*}
 (\ref{mprl_4}) & \Longleftrightarrow &
 f'(\theta) = -\frac{m}{2}[f(\theta) - l_1]\\
 & \Longleftrightarrow &
 m =
 -\frac{2f'(\theta)}{f(\theta) - l_1}.
\end{eqnarray*}
By Equation (\ref{mprl_1}), since $l_2 > f(0)$ by the definition of $f$ :
\begin{eqnarray*}
 (\ref{mprl_1}) & \Longleftrightarrow &
 f(0) = l_1 +\frac{1}{1/2[f(\theta) - l_1]^{-1} + e^p}\\
 & \Longleftrightarrow &
 p =\log\left[\frac{1}{f(0) - l_1} -\frac{1}{2[f(\theta) - l_1]}\right].
\end{eqnarray*}
By Equation (\ref{mprl_2}) :
\begin{eqnarray*}
 (\ref{mprl_2}) & \Longleftrightarrow &
 \theta =
 -\frac{1}{m}\log\left[2\frac{f(\theta) - l_1}{f(0) - l_1} - 1\right]\\
 & \Longleftrightarrow &
 \log\left[
 2\frac{f(\theta) - f(0)}{f(0) - l_1} + 1\right] -
 \frac{2\theta f'(\theta)}{f(\theta) - l_1} = 0.
\end{eqnarray*}
That achieves the proof. $\square$
\\
\\
\textit{Proof of Proposition \ref{logistic_ODE}.} For every $x\in\mathbb R$,
\begin{eqnarray*}
 f'(x) & = &
 -\frac{me^{mx + p}}{[(l_2 - l_1)^{-1} + e^{mx + p}]^2}\\
 & = &
 -m\left[l_1 +\frac{1}{(l_2 - l_1)^{-1} + e^{mx + p}} - l_1\right]\times\\
 & & \left[
 1 -\frac{1}{l_2 - l_1}\left[l_1 +\frac{1}{(l_2 - l_1)^{-1} + e^{mx + p}} - l_1\right]\right]\\
 & = &
 -m[f(x) - l_1]\left[1 -\frac{1}{l_2 - l_1}[f(x) - l_1]\right].
\end{eqnarray*}
So, the logistic function $f$ is the solution of Equation (\ref{logistic_equation}). $\square$
%


%

%

\begin{thebibliography}{99}
 \bibitem{ABM08} P. Armitage, G. Berry and J.N.S. Matthews. \textit{Statistical Methods in Medical Research.} John Wiley and Sons, 2008.
 \bibitem{AZZALINI85} A. Azzalini. \textit{A Class of Distributions which Includes the Normal Ones.} Scandinavian Journal of Statistics 12, 171-178, 1985.
 \bibitem{BGK12} T. Baghfalaki, M. Ganjali and M. Khounsiavash. \textit{A Non-Random Dropout Model for Analyzing Longitudinal Skew-Normal Response.} JIRSS 11(2), 101-129, 2012.
 \bibitem{BROWN78} C.C. Brown. \textit{Principles of Ecotoxicology (chapter 6).} John Wiley and Sons, 1978.
 \bibitem{CGN04} J.T. Chen, A.K. Gupta and T.T. Nguyen. \textit{The Density of the Skew Normal Sample Mean and its Applications.} Journal of Statistical Computation and Simulation 74, 7, 2004.
 \bibitem{GM80} R.F. Gunst and R.L. Mason. \textit{Regression Analysis and its Application: a Data-Oriented Approach.} CRC Press, 1980.
 \bibitem{HS81} N.H.G. Holford and L.B. Sheiner. \textit{Understanding the Dose-Effect Relationship : Clinical Application of Pharmacokinetic-Pharmacodynamic Models.} Clinical Pharmacokinetics 6, 429-453, 1981.
 \bibitem{LJ10} F. Lavergne and T.M. Jay. \textit{A New Strategy for Antidepressant Prescription.} Frontiers in Neuroscience, doi:10.3389/fnins.2010.00192, 2010.
 \bibitem{MARIE14} N. Marie. \textit{A Pathwise Fractional One Compartment Intra-Veinous Bolus Model.} International Journal of Statistics and Probability, doi:10.5539/ijsp.v3n3p65, 2014.
 \bibitem{PRENTICE76} R.L. Prentice. \textit{A Generalization of the Probit and Logit Methods for Dose Response Curves.} Biometrics 32, 761-768, 1976.
 \bibitem{VERLATO96} G. Verlato et al. \textit{Evaluation of Methacholine Dose-Response Curves by Linear and Exponential Mathematical Models : Goodness-of-fit and Validity of Extrapolation.} Eur. Respir. J. 9, 506-511, 1996.
 \bibitem{WAGNER07} L.A. Wagner. \textit{Some Skew Models for Quantal Response Analysis.} Ph.D. thesis, 2007.
\end{thebibliography}
\end{document}